\documentstyle[preprint,aps,epsfig]{revtex}  

\tighten       

\begin{document}

\draft
\preprint{Version of \today }

\title{Asymptotics of the trap-dominated Gunn effect in p-type Ge 
	 }

\author{L.~L.\ Bonilla$^1$ \cite{bonilla:email}, P.~J.\ Hernando$^1$ 
\cite{pjh:email}, M.~A.\ Herrero$^2$ \cite{herrero:email}, 
M.\ Kindelan$^1$ \cite{kindelan:email}, and
J.~J.~L.\ Vel\'azquez$^2$ \cite{juanjo:email} }

\address{$^1$Universidad Carlos III de Madrid, Escuela Polit\'{e}cnica 
Superior, 28911 Legan\'{e}s, Spain \\
$^2$Departamento de Matem\'atica Aplicada, Universidad Complutense 
de Madrid, 28040 Madrid, Spain}


\maketitle

\begin{abstract}
We present an asymptotic analysis of the Gunn effect in a drift-diffusion
model---including electric-field-dependent generation-recombination
processes---for long samples of strongly compensated p-type Ge at low temperature and under
dc voltage bias.  During each Gunn oscillation, there are different stages 
corresponding to the generation, motion and annihilation of solitary waves.
Each stage may be described by one evolution equation for
only one degree of freedom (the current density), except for the generation
of each new wave. The wave generation is a faster process that may be
described by solving a semiinfinite canonical problem. As a result of
our study we have found that (depending on the boundary condition) one or
several solitary waves may be shed during each period of the oscillation.
Examples of numerical simulations validating our analysis are included.

\end{abstract}

\narrowtext

\section{Introduction}
\label{intro}

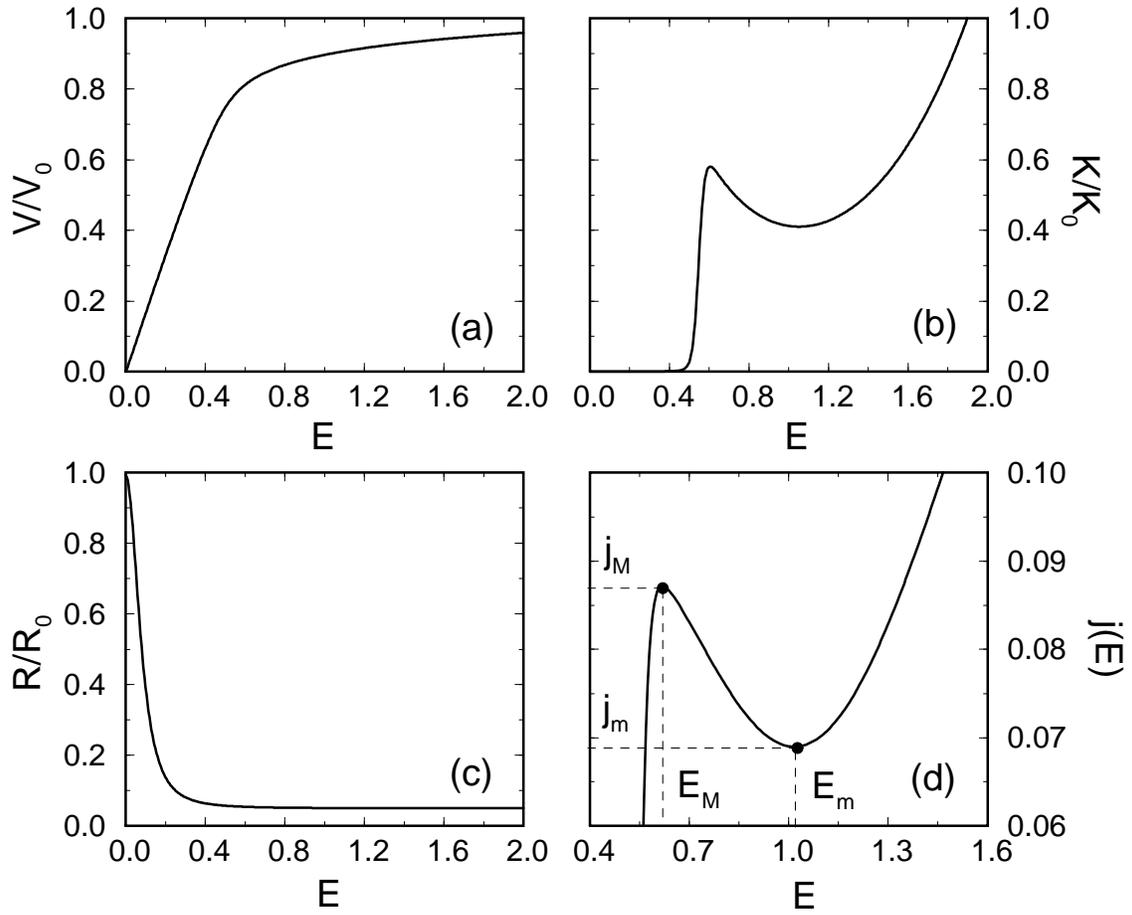
\begin{figure}   
\caption{(a) Drift velocity, (b) impact ionization coefficient, (c) recombination
coefficient, and (d) homogeneous stationary current density $j(E)$ for
$\alpha = 1.21$, illustrating the NDR in our model. }
\label{j_sh}
\end{figure}

In recent years a great variety of oscillatory behaviors have been observed
in semiconductors displaying nonlinear electrical conduction. A particularly
interesting system to study spatiotemporal phenomena in the laboratory
is ultrapure extrinsic cooled bulk p-type Ge under voltage bias conditions, 
\cite{teitsPRL83,kahnPRB91}. Observed phenomena include time-periodic 
oscillations of the current on a purely resistive external circuit under dc 
voltage bias due to the periodic creation of a solitary wave at the 
injecting contact, its motion inside the semiconductor and its annihilation 
at the receiving contact \cite{kahnPRB91}. Essentially this is the same as 
the usual Gunn effect in n-GaAs \cite{gunIJRD64}, except that the solitary
wave in p-Ge moves much more slowly than the carriers (the case in the usual
Gunn effect) due to the generation-recombination dynamics of ionized traps 
which dominate the transport properties. Other experimental observations 
include intermittency near the onset of the oscillatory instability 
\cite{kahnPRB92,kahn45PRB92},
and spatiotemporal chaos under dc$+$ac voltage bias \cite{kahnPRL92}.
These phenomena have been studied theoretically by means of a drift-diffusion
model which includes impurity trapping of mobile holes and impact ionization
of neutral acceptors \cite{teitsPRL84,wesJAP85}. In dimensionless form,
the model equations are \cite{bonPRB92}:
\begin{eqnarray}
\frac{\partial A}{\partial t} = \frac{\Gamma}{\beta}\, (\alpha - 1 -
A) + P\left [(\alpha-1) K(E)\right. \nonumber \\
\left. \mbox{} - R(E) - [K(E)+R(E)]\, A \right],
\label{1}\\
\beta \frac{\partial E}{\partial t} = J - V(E) P + \delta
\frac{\partial P}{\partial x},
\label{2}\\
\frac{\partial E}{\partial x} = P-A,
\label{3}\\
{1\over L}\,\int_0^L E(x,t)\ dx = \phi. \label{4}
\end{eqnarray}
The unknowns in these equations are the electric field $E(x,t)$, the hole
concentration $P(x,t)$, the concentration of the ionized acceptors $A(x,t)$ 
and the total current density $J(t)$. $\phi$ is the constant dc 
bias. To solve this system of equations, we need
two boundary conditions (representing the effect of the contacts at $x=0$ and
$x=L$) and two initial conditions. In Section \ref{model} we shall prove that
we have a well-posed problem with positive solutions for all time for the 
following Dirichlet boundary conditions:
\begin{equation}
P(0,t) = P_0,\quad\quad P(L,t) = P_1 . \label{bcD}
\end{equation}
(the $P_i$'s are positive numbers). These conditions have been
considered before in \cite{haeIP89}. A discussion of imperfect boundary 
conditions for the Gunn effect introducing the concept of contact 
characteristic was first published by H.\ Kroemer \cite{kro68}. Ohmic boundary 
conditions in
which the field at the contact is proportional to the current density may give
rise to negative current densities and numerical instabilities as time elapses.
Eq.\ (\ref{1}) is a rate equation for the  ionized acceptors including the 
processes of photogeneration, impact ionization and recombination 
(coefficients $\Gamma$, $K(E)$ and $R(E)$, respectively). Eq.\ (\ref{2}) is 
a form of Amp\`ere's law saying that the displacement current is equal to 
the total current density minus the drift-diffusion current due to the free 
holes. (\ref{3}) is the Poisson equation and Eq.\ (\ref{4}) is the voltage 
bias condition. The transport coefficients $V(E)$, $K(E)$ and $R(E)$ depend 
nonlinearly on the electric field as shown in Fig.~\ref{j_sh} \cite{wesJAP85}. 
The qualitative nature of much of the predicted behavior---e.g., the
instability of the stationary electric-field profile and the stability of
propagating high-field domains---depends only on the presence of a negative
slope of the homogeneous stationary current density $j(E)$ (defined below)
over an interval of positive fields,
\cite{bonPRB92,bonSST94} and not on the exact form of the underlying
coefficients.  Other authors have stated different forms of the coefficient
curves for various reasons.\cite{mitAPA86,schSSE88,schAPA89} Recently, Monte
Carlo simulation has been used to provide a more rigorous determination of the
coefficients for a related p-type system,\cite{kuhPRB93} however such
simulations do not yet exist for the closely compensated Ge samples under 
consideration in this paper.
The homogeneous stationary current density $j(E)$ is given by the formula
\begin{equation}
j(E) = \left( \frac{\alpha K(E)}{K(E)+R(E)} - 1\right)\, V(E), \label{5}
\end{equation}
\cite{bonPRB92,canPRB93,bergPRB96}. If the compensation ratio $\alpha$ (the
ratio of the acceptor concentration to the donor concentration) is only 
slightly larger than 1 (see \cite{teits94} for the details), $j(E)$ is
N-shaped for large enough positive fields. Then there is an interval
$(E_M,E_m)$ between the abcissas of the maximum [$j(E_M) = j_M$] and the 
minimum [$j(E_m) = j_m > 0$] of $j(E)$ for which $dj/dE<0$ and $j(E)>0$.
See Fig.~\ref{j_sh}.

 The dimensionless parameters
$\beta$, $\delta$, and $\Gamma/\beta$ are very small for the p-Ge and can be
set to zero in the leading order approximation \cite{bonPRB92}. Then $P=J/V(E)$ 
from (\ref{2}), and (\ref{1}) and (\ref{3}) become: 
\begin{eqnarray}
\frac{\partial A}{\partial t} & = & {J\, [K(E)+R(E)]\over V(E)^{2}}\, [j(E) -
V(E)\, A] , \label{1bis} \\
\frac{\partial E}{\partial x} & = & {J\over V(E)} - A.
\label{3bis}
\end{eqnarray}
Using Eq.\ (\ref{3bis}), we may eliminate $A$ in favor of $E$ and then insert
the result into (\ref{1bis}). We obtain the following reduced equation
\cite{bonPD91,bonPRB92,bonPD92}:
\begin{eqnarray}
\frac{\partial^{2}{E}}{\partial{x} \partial{t}} + J\,
\frac{K(E)+R(E)}{V(E)^{2}}\,\left(\frac{V'(E)}{K(E)+R(E)}\, 
\frac{\partial E}{\partial t} \right.\nonumber\\
\left. + V(E)\, 
\frac{\partial E}{\partial x} + j(E) - J\right) = \frac{1}{V(E)}\,\frac{dJ}{dt},
\label{E}
\end{eqnarray}

The reduced model equations (\ref{E}) and (\ref{4}) for $E(x,t)$ and $J(t)$
are supplemented with the boundary condition at the injecting contact $J(t)
= P_0 \, V[E(0,t)]$. According to this equation, the electric field at the 
contact is a nonlinear function of the current. In \cite{bonPD92} it was shown
that the resulting stable current oscillations are qualitatively like the
solutions of the same reduced problem with the linear relation:
\begin{equation}
E(0,t) = \rho J(t), \label{bc}
\end{equation}
(Ohm's law; $\rho>0$ is the dimensionless contact resistivity) and an initial 
condition $E(x,0)$. Although much work has been done on the reduced model 
problem (\ref{E}), (\ref{4}) and (\ref{bc}) [see \cite{bergPRB96} 
and references therein; equivalently we may analyze (\ref{1bis}), (\ref{3bis}),
(\ref{4}) and (\ref{bc})], important basic questions concerning the asymptotic
description of their stable solutions are still open. In this paper we present
an asymptotic description of the Gunn effect in p-Ge and compare the results 
with numerical simulations. 

The paper is organized as follows.  Section~\ref{model} contains a proof of the
global existence of positive solutions to the full drift-diffusion model with 
particular boundary conditions and a brief review of the known results for the 
reduced model equations.  In Sec.~\ref{asymptotics}, we introduce our
asymptotic scaling, derive the outer and inner solutions, and put the
pieces together thereby describing one period of the solution through the 
succession of its different stages. The delicate problem of shedding new waves 
through an instability of the boundary layer near the injecting contact is 
discussed in Section \ref{firing}. Sec.~\ref{conclusion} contains our 
conclusions and some open problems. Appendix A contains technical details on 
the shedding problem.

\section{Model system and global existence and uniqueness of solutions}
\label{model}

We shall next show that, under rather mild assumptions on coefficients
and data, the system (\ref{1})-(\ref{5}) has a unique global solution when
supplemented with initial conditions 
\begin{eqnarray}
A(x,0) \, = \, A_0(x), \qquad P(x,0) \, = \, \overline{P}(x).
\label{j9}
\end{eqnarray}
Specifically, we will require functions $A_0$, $\overline{P}$, $P_0$ 
and $P_1$ ($P_0$ and $P_1$ may be known functions of time, not just
constants) to be bounded, whereas functions $K(E)$, $R(E)$ and
$V(E)$ will be assumed to be $C^1$. Furthermore, we shall impose
the following growth condition on $K(E)$:
\begin{eqnarray}
\left| K(E) \right| \, \leq \, C \left( \left| E \right|^{\gamma} \, + 
\, 1 \right) \quad \mbox{ for some } \gamma >0
\label{j10} 
\end{eqnarray}
(which is perfectly reasonable in view of the physics behind the model; see
\cite{bergPRB96,teits94}).

Without loss of generality, we may assume $P_0 = P_1 = 0$ in
(\ref{5}), and $\Gamma = 0$ in (\ref{1}). As a matter of fact, the term
thus dropped in (\ref{1}) is a linear one and therefore harmless from
the point of view of global existence. For the sake of notational simplicity,
we shall also write:
\begin{eqnarray}
f(A,E) = (\alpha-1) K(E) - R(E) - [K(E)+R(E)]\, A. \label{j11} 
\end{eqnarray} 
Differentiating (\ref{2}) with respect to the space coordinate $x$
and using (\ref{3}), we find the equation of charge continuity:
\begin{eqnarray}
\beta \, P_t \, = \, \delta P_{xx} \, - \left( V(E) \, P \right)_x \,
+ \, \beta \, A_t ,\label{j17}   
\end{eqnarray}
which is equivalent to 
\begin{eqnarray}
\beta \, P_t \, = \, \delta P_{xx} \, - \left( V(E) \, P \right)_x \,
+ \, \beta \, f(A,E) \, P,
\label{j12}
\end{eqnarray}
once (\ref{1}) is substituted into (\ref{j17}).
Let us discuss first local existence of solutions. To this end, we observe
that on setting $\psi = - \int_0^x E(\xi,t)d\xi$ (the electric potential), 
equations (\ref{3}) and (\ref{5}) read:
\begin{eqnarray*}
-\psi_{xx} \, = \, P \, - \, A; 
\qquad \psi(0,t) \, = \, 0, \quad \psi(L,t) \, = \, -\phi \, L
\end{eqnarray*}
By classical results for Poisson's equation, we then deduce that:

\bigskip

\noindent {\bf Lemma 1.} {\em If $P$ and $A$ are bounded, then $\left| E_x 
\right| \, = \,  \left| \psi_{xx} \right| \, \leq \, C$ for any
$t \, > \, 0$, where $C$ depends only on the bounds for $P$ and $A$.}

\bigskip

\noindent On the other hand, integrating in (\ref{3}) gives:
\begin{eqnarray*}
E \, = \, \int_0^x \, \left[
P(\xi,t) \, - \, A(\xi,t) \right] \, d\xi \, + \, C_1,
\end{eqnarray*}
where $C_1 \, = \, C_1(\phi, P, A)$. Hence, equation (\ref{j12}) may be 
rewritten in the form:
\begin{eqnarray}
\beta \, P_t \, = \, \delta P_{xx} \, - \left( V(E) \, P \right)_x \,
+ \, \beta \, f(A,E(P,A))\, P \label{j14} \\
\qquad \equiv \, \delta P_{xx} \, + \,
{\cal M} 
(P, P_x, E, E_x,A), \nonumber
\end{eqnarray}
where the operator 
${\cal M}$ is locally Lipschitz in its arguments. To proceed
further, we remark that an integration in time of (\ref{1}) yields
\begin{eqnarray}
A(x,t) = A(x,0) + \int_0^t 
f[A(x,s),E(P,A)]\, P(x,s) ds
\label{j15}
\end{eqnarray}
In view of (\ref{j14}) and  (\ref{j15}), local existence follows now by
means of a contractive fixed point argument for the following operator:
\begin{equation}
T \left( 
\begin{array}{c}
P \\ \\A 
\end{array} 
\right) =
\left( 
\begin{array}{l}
\frac{\delta}{\beta} 
S(t) \overline{P} +
\frac{1}{\beta} \int_0^t S(t-s) {\cal M}
(P,P_x,E,E_x,A) ds \\ \\
A(x,0) \, + \, \int_0^t f(A(x,s), E(P,A) )
\end{array}
\right)
\label{j16}
\end{equation}
where $S(t)$ denotes the heat semigroup in $[0,L]$ with
homogeneous Dirichlet conditions. The operator $T$ is to be considered 
as acting on the space of functions $(P,A)$ such that
$P\in L^{\infty}(0,t^*;H^1(0,L))$ and $A \in
L^{\infty}(0,t^*;L^{\infty}(0,L))$, where $t^*$ is a sufficiently 
small positive number. On using the standard estimate
\begin{eqnarray*}
\| S(t) \, P \|_{H^1(0,L)} \, \leq \, 
C \, t^{-1/2} \, \| P \|_{L^2(0,L)} 
\end{eqnarray*}
for some $C > 0$ (see for instance \cite{henry}), we readily derive 
the contractivity of the operator $T$ for $0<t^*\ll 1$. This gives 
at once local existence and uniqueness.
We next set out to describe our global existence argument. To that end,
we shall first prove that the following result holds for (\ref{j17}):

\bigskip

\noindent {\bf Lemma 2.} {\em Assume that $A(x,t)$ and $\overline{P}(x)$ 
are bounded. Then for any $\alpha_{1}>1$ there exists $C=C(\alpha_{1})>0$ 
such that}
\begin{eqnarray}
\left(\int_0^t \, \int_0^L \,
\left| P(x,t) \right|^{\alpha_{1}} \, dx \, dt \right)^{1/\alpha_{1}} 
\nonumber\\ 
\leq \, C \, \left( \| A \|_{\infty} \, (t+1) \, + \, 
\| \overline{P} \|_{\infty} \right), \label{j18}
\end{eqnarray}
{\em for any $t>0$ for which a solution of our problem exists.}

\bigskip

We remark in passing that we are denoting by $\| A \|_{\infty}$
a bound on the supremum of $A$. To prove Lemma 2, we argue as 
follows. Set
\begin{eqnarray}
u\, = \, \int_0^t \, P(x,s) \, ds
\end{eqnarray}
Then $u$ satisfies
\begin{eqnarray}
\beta \, {\partial u\over\partial t} & = & \delta\, {\partial^{2} u\over\partial 
x^{2}} + \beta \, A(x,t) \nonumber\\
& - & \frac{\partial}{\partial x} \left( \,
\int_0^t \, V(E(x,s)) \, P(x,s) \, ds \, \right).
\label{j19}   
\end{eqnarray}
If the last terms on the right of (\ref{j16}) were absent, standard 
parabolic theory would yield at once
\begin{eqnarray}
\left( \, \int \int_Q \, |u_t|^{\alpha_{1}} \, dx \, dt 
\right)^{1/\alpha_{1}}\, \leq \,
C \, \left( \| A \|_{\infty} \,  + \, 
\| \overline{P} \|_{\infty} \right)
\label{j20} \\  
\quad \mbox{for any } \alpha_{1}>1,\nonumber
\end{eqnarray}
where $Q=(0,L) \times (0,t^*)$ and $C=C(L,t^*)>0$. That would yield
(\ref{j18}) except for the linear dependence on $t^*$ stated there.
To prove Lemma 2 in the general case, we shall assume for
simplicity that $\overline{P} \equiv 0$, and note that a solution of
(\ref{j19}) with homogeneous initial and side conditions may be 
represented in the form
\begin{eqnarray*}
P(x,t) = \int_0^t S(t-s) \, \left\{ A_t  -
\frac{1}{\beta}\, \left(
V(E)  P \right)_x \right\}\, ds \\
\qquad \equiv \, P_1(x,t) \, + \, P_2(x,t)
\end{eqnarray*}
Recalling (\ref{j20}), we see that any bound for 
$\|~A~\|_{\infty}$ would yield at once a bound for
$\|~P_1(x,t)~\|_{\infty}$. On the other hand,
if $\|~P~\|_{L^{\alpha_{1}}}$ is bounded, then 
$(VP)_x \, \in W^{-1,\alpha_{1}}$ for
$\alpha_{1} \gg 1$, whence $|\int_0^t S(t-s) (VP)_x\, ds|\le C\, 
t^\nu\, \|~VP~\|_{\infty}$ for some positive $C$ and $\nu$. A fixed 
point Theorem as that sketched 
to obtain local existence will then give (\ref{j18}) for
$0<t<\delta_1 \ll 1$. On iterating in time the previous argument,
(\ref{j18}) now follows for the general case thus proving Lemma 2. 

We next observe that, by Lemma 1 and (\ref{j18})
\begin{eqnarray}
\left(\int_T^{T+1} \, \left( \, \sup_{(0,L)} |E(x,t)|^{\alpha_{1}} 
\right)\,  dt \right)^{1/\alpha_{1}} \, \leq \,
C \, \left( 1 \, + \, T \right).
\label{j21}
\end{eqnarray}
where $C=C(\| A \|_{\infty}, \| A \|_{\infty})$.

Taking into account the growth condition (\ref{bc}) it then turns out
that $|f(A,E)| \, \leq \, C(|E|^{\gamma} + 1)$ for some
$C=C(\|~A~\|_{\infty})>0$. By (\ref{j21}), we now see that
\begin{eqnarray}
\left(\int_T^{T+1} \, \int_0^{L} 
\, |f(A,E)|^{\beta_{1}} \,  dx \, dt \right)^{1/\beta_{1}} \, \leq \,
C \, \left( 1 \, + \, T \right) \, ,
\label{j22}
\end{eqnarray} 
where $\beta_{1}\gg 1$ and $C=C(\beta_{1})>0$.

Putting together (\ref{j22}) and (\ref{j18}), we see that $\sigma =
\sigma(\alpha_{1},\beta_{1}) > 1$ exists such that
\begin{eqnarray}
\left(\int_T^{T+1} \, \int_0^{L} 
\, |f(A,E) \, P|^{\sigma} \,  dx \, dt \right)^{1/\sigma} \, \leq \,
C \, \left( 1 \, + \, T \right)^{\theta}.
\label{j23}
\end{eqnarray} 
where $\theta= \theta(\sigma)>1$. In view of (\ref{1}), $A_t$ also
satisfies (\ref{j21}), and this in turn implies that
\begin{eqnarray}
|P(x,t)| \, \leq \, C \left( 1 \, + \, t \right)^{\theta_1} \qquad
\mbox{for some } \theta_1>0,
\end{eqnarray} 
 whereas
\begin{eqnarray}
|A(x,t)| \, \leq \, C \left( 1 \, + \, t \right)^{\theta_2} \qquad
\mbox{for some } \theta_2>0,
\end{eqnarray}
both bounds being valid as long as $P$ and $A$ exist. This gives at
once global existence.

\section{Asymptotic analysis}
\label{asymptotics}
It is convenient to redefine the time and space scales in such a way that the
semiconductor length becomes 1:
\begin{equation}
\epsilon = {1\over L}\, ,\quad y =  {x\over L}\, ,\quad s = {t\over L}\, . 	
\label{a1}
\end{equation}
Then Eqs.~(\ref{4})-(\ref{bc}) become
\begin{eqnarray}
J - j(E) = \epsilon \left(\frac{1}{K(E)+R(E)}\,\left[ V'(E)\,
\frac{\partial E}{\partial s} \right.\right.\nonumber\\
\left.\left.- V(E)\, \frac{d\ln J}{ds}\right]
 + V(E)\, \frac{\partial E}{\partial y} \right)\nonumber\\
+ \epsilon ^2 \, \frac{V(E)^{2}}{J\, [K(E)+R(E)]}\,
\frac{\partial^{2}{E}}{\partial{y} \partial{s}}\, ,
\label{a2}
\end{eqnarray}

\begin{eqnarray}
\int_{0}^{1} E(y,s) \,dy = \phi,														       \label{a3}\\
E(0,s) = \rho J(s), 																						\label{a4}
\end{eqnarray}

\noindent in terms of the `slow' variables $s = \epsilon t$ and $y = \epsilon x$. 
We shall describe the time-periodic solution of these equations in the limit
$\epsilon \to 0$ ($L\to \infty$) by leading order matched asymptotic expansions.

\subsection{Outer solutions}
Clearly the leading order of the outer expansion of the solutions to
Eqs.\ (\ref{a2}) - (\ref{a3}) yields 
\begin{equation}
J - j(E) =  0.														       \label{a5}
\end{equation}
Then the outer electric field is a piecewise constant function whose profile
is a succession of zeros of $J - j(E)$, $E_k(J)$ ($k = 1,2,3$) separated
by discontinuities. These discontinuities are traveling wavefronts whose
velocities can be found by studying the inner solutions. Typically pairs of
these wavefronts bounding a region where the electric field is $E_3(J)$ 
constitute solitary waves, which are the basis of our study. See below.

\subsection{Inner solutions}
Near $y=0$ or near the moving discontinuities in the outer field profile, there
are regions of fast variations of the electric field. In them $J=J(s)$, 
$E\sim F(x,t;s)$, where $x=y/\epsilon$ and $t=s/\epsilon$ according to 
Eq.\ (\ref{a1}). 

\subsubsection{Injecting layer}
The field in the injecting layer (defined as the boundary layer near $y=0$) 
obeys the equations:
\begin{eqnarray}
\frac{V(F)^{2}}{J(s)\, [K(F)+R(F)]}\,\left(\frac{\partial^{2}{F}}{\partial{x} 
\partial{t}} + \frac{J\, V'(F)}{V(F)^{2}}\,\frac{\partial F}{\partial t}\right) 
\nonumber\\
 + V(F)\, \frac{\partial F}{\partial x} = J(s) - j(F) \nonumber\\
 + \epsilon \frac{V(F)}{K(F)+R(F)}\, \frac{d\ln J}{ds} ,\label{a6}
\end{eqnarray}
\begin{eqnarray}
F(0,t;s) = \rho\, J(s). \label{a7}
\end{eqnarray}
Except in very short time intervals where a new wave is being formed,
$F=F(x;s)$ is a quasistationary monotonically decreasing profile joining
$F=\rho J(s)$ and $F=E_1[J(s)]$, which solves the equation:
\begin{equation}
V(F) \, \frac{\partial F}{\partial x} = J(s) - j(F) .\label{a8}
\end{equation}

\subsubsection{Traveling wavefronts}
Near a discontinuity of the outer electric field profile, located at e.g.
$y=Y(s)$, the electric field is a traveling wavefront solution of (\ref{a6}), 
$F=F(\chi;s)$, with $\chi = [y - Y(s)]/\epsilon$ and $dY/ds = c$:
\begin{eqnarray}
c \, \frac{\partial^{2}{F}}{\partial \chi^{2}} + J(s)\,
\frac{K(F)+R(F)}{V(F)^{2}}\
\left(\left[\frac{c\, V'(F)}{K(F)+R(F)} \right.\right. \nonumber\\
\left.\left. - V(F)\right]\, 
\frac{\partial F}{\partial \chi} + J(s) - j(F) \right) = 0\, .
\label{a9}
\end{eqnarray}

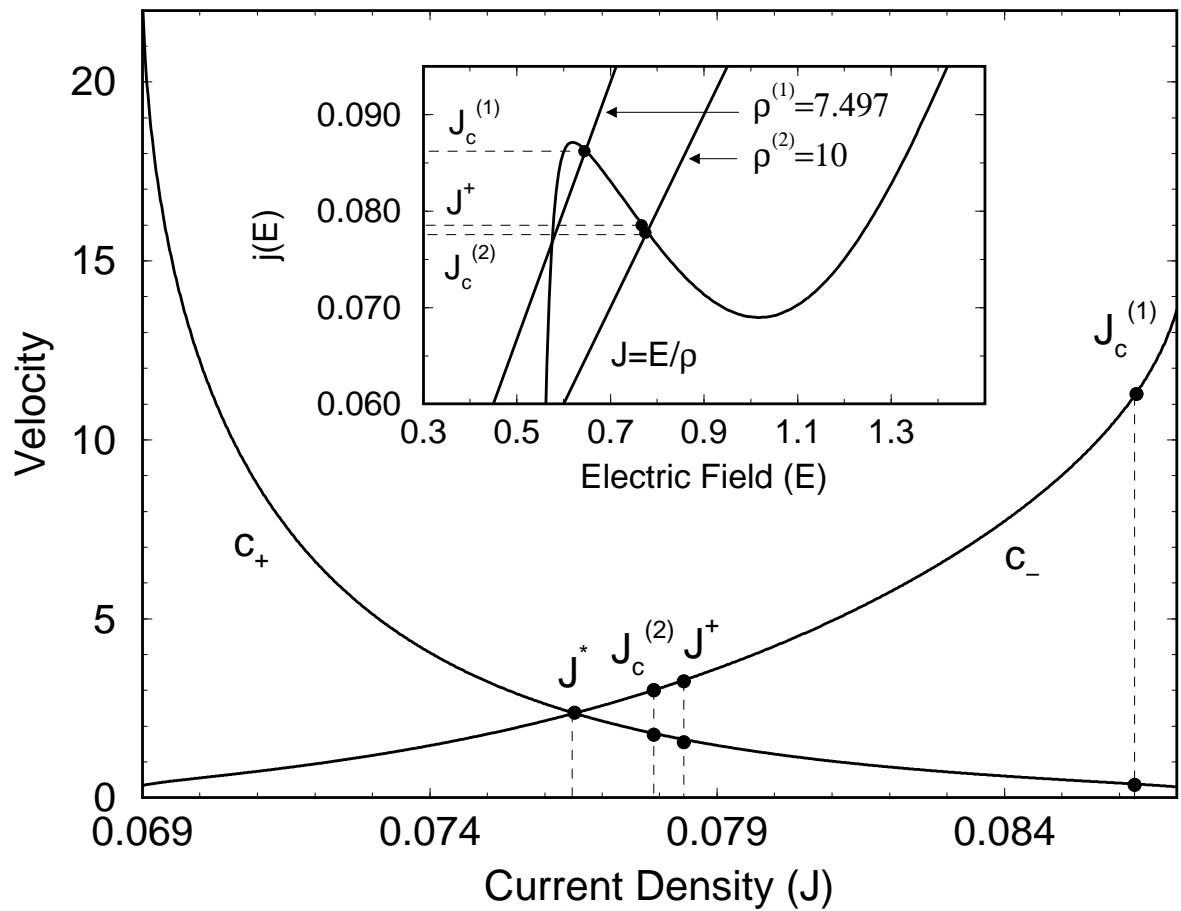
\begin{figure}   
\caption{Velocities $c_+$ and $c_-$ of the heteroclinic wavefronts as functions 
of the current density. Notice that the lines intersect at $J=J^*\approx
0.076515$. We have also
marked the current $J^{\dag}\approx 0.07840$ at which $2c_+ = c_-$.
$J_{c}^{(1)}$ and  $J_{c}^{(2)}$ are the critical currents $J_{c}$
for the resistivities of Figures 3 and 4, respectively. The inset shows
the location of $J^{\dag}$, $J_{c}^{(1)}$ and  $J_{c}^{(2)}$ on 
the bulk current density curve $j(E)$.
}
\label{c(J)}
\end{figure}

These wavefronts should join the two different stable zeros of $j(E)-J$,
$E_1(J)$ and $E_3(J)$. To find them we have to (i) fix $J(s)=J$, (ii) vary
$c(J)$ until we find a solution of (\ref{a9}) which is equal to $E_1(J)$
[resp. $E_3(J)$] when $\chi = -\infty$ and to $E_3(J)$ [resp. $E_1(J)$]
when $\chi = +\infty$. Thus there are two types of wavefronts, each with a 
given wavespeed $c$ , which is a unique function of $J(s)$: 
\begin{itemize}
\item A heteroclinic solution of the phase plane corresponding to (\ref{a9}) 
joining the saddles $(E_1(J),0)$ and $(E_3(J),0)$ with $\partial F/\partial
\chi >0$. For each $J\in (j_m,j_M)$ there is one such solution with wavespeed 
$c_+(J)$.
\item A heteroclinic solution of the phase plane corresponding to (\ref{a9}) 
joining the saddles $(E_3(J),0)$ and $(E_1(J),0)$ with $\partial F/\partial
\chi <0$. For each $J\in (j_m,j_M)$ there is one such solution with wavespeed 
$c_-(J)$.
\end{itemize}
The functions $c_+(J)$ and $c_-(J)$ are depicted in Fig.~\ref{c(J)}. Notice 
that they cross at a value $J=J^\ast$ which was already defined in
\cite{bonPD91} as the value of the current for which a homoclinic orbit
joining $(E_1,0)$ to itself `collides' with the other saddle $(E_3,0)$.

\subsection{Putting the pieces together}
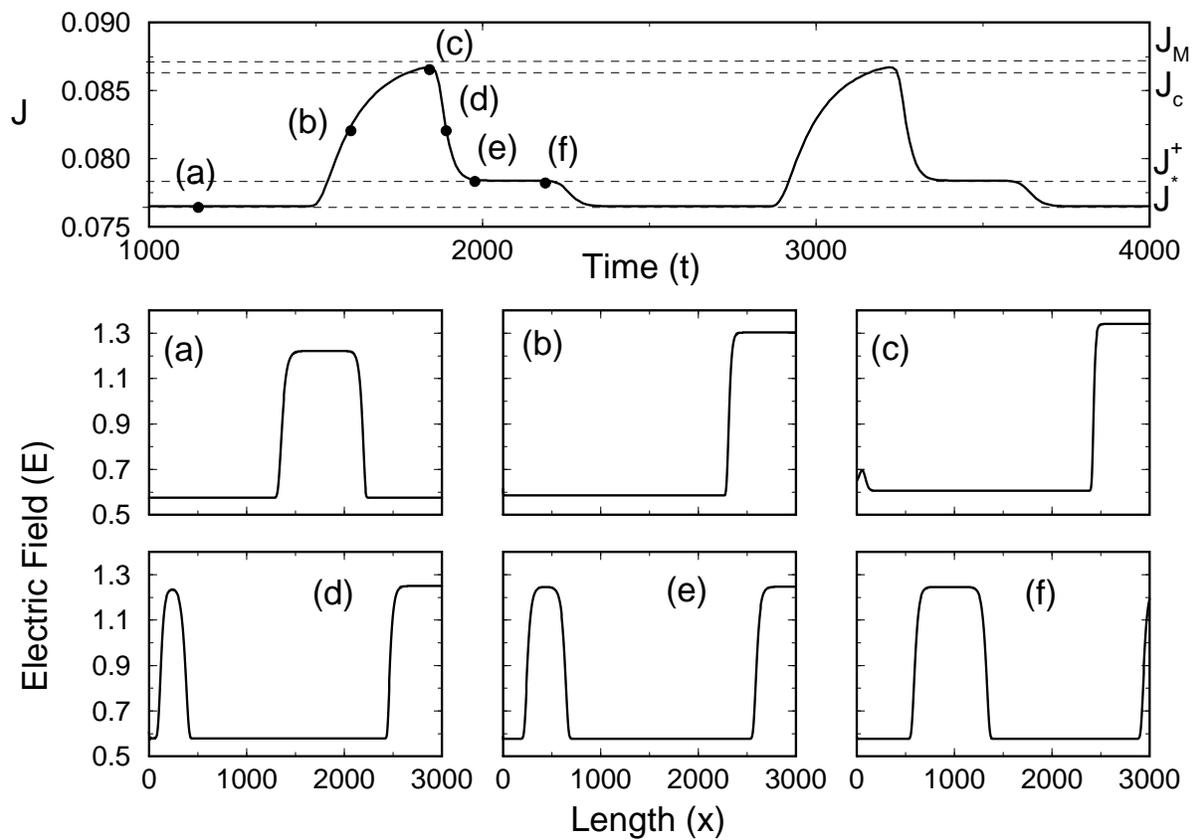
\begin{figure}   
\caption{ Numerical solution of the reduced model during one period of the 
oscillation in the stable case $2c_+(J_c) < c_-(J_c)$ for which only one 
new wave per period is shed. Parameter values are $L=3000$, $\rho = 7.49736$,
$\phi = 0.749565$. }
\label{simulation1}
\end{figure}

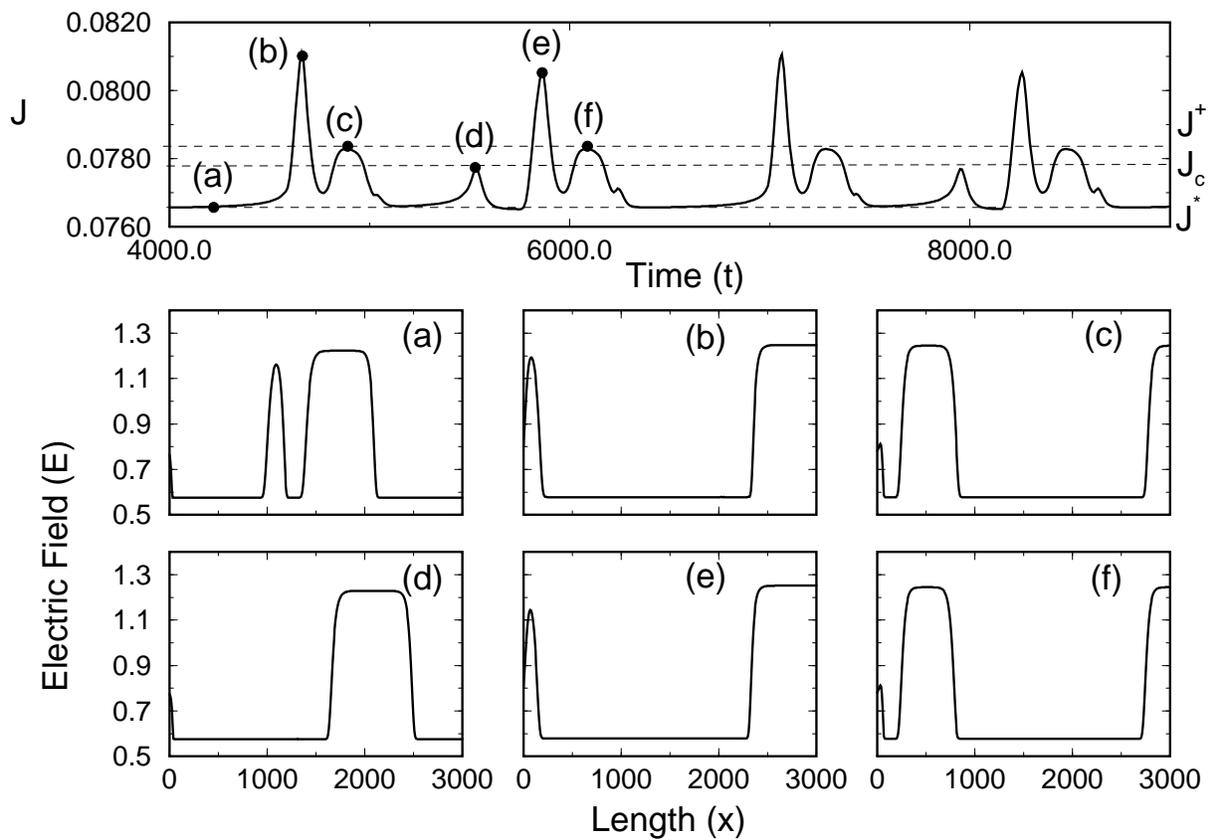
\begin{figure}   
\caption{ Numerical solution of the reduced model during one period of the 
oscillation in the unstable case $2c_+(J_c) > c_-(J_c)$, corresponding to
$\rho = 10$ and the rest of parameters as in Figure 3. With these values of
the parameters, two new waves are shed during each period.}
\label{simulation2}
\end{figure}

We shall start at a time when there is only one solitary wave in the sample,
far from its ends. We shall assume that $E_M<\phi<E_m$ in Fig.~\ref{j_sh}
and that $E_M/j_M<\rho < E_m/j_m$. Then a Gunn effect mediated by solitary
waves occurs \cite{bonPD92}. The initial outer field profile is
\begin{eqnarray}
E(y,0) = E_1(J) + [E_3(J)-E_1(J)]\, [\theta(Y_2-y)\nonumber\\
 - \theta(Y_1-y)] + O(\epsilon), \label{a10}\\
Y_2 - Y_1 = \frac{\phi - E_1(J)}{E_3(J) - E_1(J)}\, .\label{a11}
\end{eqnarray}
Here $Y_1$ and $Y_2$, $Y_1<Y_2$ are the positions of the fronts connecting 
$E_1(J)$ and $E_3(J)$. Fixing the values of $\phi$ and $J(0)$, (\ref{a11}) 
fixes $Y_2(0) - Y_1(0)$. $\theta(x)$ is the unit step function: $\theta(x)=1$ 
if $x>0$, $\theta(x)=0$ if $x<0$. The outer field profile for $s>0$ is given
by (\ref{a10}) with $J=J(s)$, and $Y_i=Y_i(s)$ ($i=1,2$). The location of
the wavefronts is given by the equations:
\begin{equation}
{dY_{1}\over ds} = 	c_+(J),\quad {dY_{2}\over ds} = 	c_-(J),		\label{a12}
\end{equation}
whereas their separation is related to the bias $\phi$ through (\ref{a11}).
We can find an equation for $J$ by differentiating (\ref{a11}) and then 
inserting (\ref{a12}) in the result. We obtain:
\begin{equation}
{dJ\over ds} = 	\frac{(E_{3}-E_{1})^{2}}{{\phi-E_{1}\over j'_{3}} +{E_{3}-\phi
\over j'_{1}}}\, (c_+ -	c_-)\,		\label{a13}
\end{equation}
where $j'_iÊ\equiv j'[E_i(J)]$. This is a simple closed equation for $J$ that 
demonstrates that $J$ tends to $J^\ast$ (for which $c_+=c_-$) exponentially fast. 
After a certain time, the wavefront
$Y_2$ reaches 1 and we have a new stage governed by (\ref{a10}) with $Y_2=1$
and $Y_1$ given by (\ref{a11}). The equation for $J$ becomes
\begin{equation}
{dJ\over ds} = 	\frac{(E_{3}-E_{1})^{2}}{{\phi-E_{1}\over j'_{3}} +{E_{3}-\phi
\over j'_{1}}}\, c_+ > 0\,		\label{a14}
\end{equation}
and $J$ increases until it surpasses the value $J_c$ such that $E_2(J)=\rho 
J$ at some time $s=s_1$. After this time the boundary layer near $y=0$ is
no longer quasistationary. In fact, about this time a new stage begins 
during which the boundary layer becomes unstable (see Section \ref{firing}) 
and it starts shedding a new solitary wave on the fast time scale $\tau 
=(s-s_1)/\epsilon$. Meanwhile the current density reaches its maximum and 
it decreases until $J=J_c$ again later on. During this stage the new 
solitary wave is generated and it starts departing from $y=0$, a process 
which will be described in the next Section. Let us call $\tau_2>0$ the 
fast time at which the solitary wave is fully formed, so that the field 
inside it is $E_3(J)$, and outside it is $E_1(J)$. The wavefronts 
enclosing the high field region are centered at $y=Y_3$ and $y=Y_4$, $Y_3
<Y_4$. We then have another slow stage, during which the outer electric 
field profile is:
\begin{eqnarray}
 \phi = E_1(J) + [E_3(J)-E_1(J)]\, (1-Y_1 + Y_4 - Y_3)\nonumber\\
 + O(\epsilon),	
\label{a15}
\end{eqnarray}
(assuming that $\phi$ is so large that the old wave $Y_1(s)<1$). Differentiating
(\ref{a15}) with respect to $s$ and using that $Y_1$ and $Y_3$ move with speed
$c_+(J)$ while $Y_2$ moves with speed $c_-(J)$, we obtain:
\begin{equation}
{dJ\over ds} = 	\frac{(E_{3}-E_{1})^{2}}{{\phi-E_{1}\over j'_{3}} +{E_{3}-\phi
\over j'_{1}}}\, (2 c_+ -	c_-)\,	.	\label{a16}
\end{equation}
We have now two possibilities: either (i) $(2 c_+ -	c_-)<0$ and then $J$ 
further decreases until the zero $J^{\dag}$ of $[2 c_+(J) -	c_-(J)]$ is reached 
(again for biases so large that the old wave at $Y_1(s)<1$); see 
Fig.~\ref{simulation1}, or (ii) $(2 c_+ -	c_-)>0$ and new wave(s) are shed; 
see Fig.~\ref{simulation2}. We consider only the stable case 
(i) which occurs for $J_c>J^{\dag}$ large enough (or equivalently, for small 
enough contact resistivity $\rho$). In this case $J(s)$ decreases 
exponentially fast towards $J^{\dag}$, thereby forming a plateau when the bias 
$\phi$ is large enough. After the old wave reaches $y=1$, we get again 
Equations (\ref{a10}) -- (\ref{a13}) with $Y_3$ and $Y_4$ playing the roles 
of the previous wavefronts $Y_1$ and $Y_2$, respectively. We thus recover 
the initial situation and a full period of the Gunn oscillation is completed. 

\section{Shedding new waves from the injecting layer}
\label{firing}
After the current density surpasses the value $J_c$ at $s=s_1$, a new stage 
begins during which a new wave is shed from the injecting layer. To understand 
how the quasistationary injecting layer solution becomes unstable and a new wave
is shed, it is convenient to keep two terms in our asymptotic outer expansion 
of the solution: the second order term in the outer expansion enters at the
same order as the leading order term of the injecting layer in the bias 
condition. It can be seen that the stability of the injecting layer is 
governed by equations similar to those of the stationary solutions 
\cite{bonSST94,bonSJAM94}, except that the old wave located at $y=Y(s_1)$ needs 
to be taken into account in the outer quasistationary profile. Then a similar 
instability criterion holds, and the injecting layer becomes unstable when 
$F\sim E_2(J_c)$ on an interval of width $O(\epsilon\ln\epsilon)$ next to $y=0$. 

      The appropriate time scale for the shedding stage is $\tau = 
(s-s_1)/\epsilon$, so that position of the old wavefront is $Y_1 \sim Y_1(s_1) +
\epsilon c_+ \tau$ during this stage. Inserting the Ansatze
\begin{eqnarray}
E = E^{(0)}(y,\tau) + \epsilon E^{(1)}(y,\tau) + O(\epsilon^2),\nonumber\\
A = A^{(0)}(y,\tau) + \epsilon A^{(1)}(y,\tau) + O(\epsilon^2),\nonumber\\
J = J_c + \epsilon J^{(1)}(\tau) + O(\epsilon^2),\label{outer2}
\end{eqnarray}
in Eqs.~(\ref{1bis}) and (\ref{3bis}), we find that the leading order term 
$E^{(0)}(y,\tau)$ is given by (\ref{a10}) with $Y_1 \sim Y_1(s_1) + \epsilon 
c_+ \tau$ and $Y_2=1$, and $J=J_c$. After some algebra, the equations for 
$E^{(1)}$ and $A^{(1)}$ turn out to be
\begin{eqnarray}
E^{(1)} = \frac{V_{i}}{J_{c} V'_{i}}\, (J^{(1)} - V_{i} A^{(1)}), \label{E_1}\\
\frac{\partial A^{(1)}}{\partial \tau} + \frac{K_{i}+R_{i}}{V'_{i}} j'_i \, 
A^{(1)} \nonumber\\
= \frac{K_{i}+R_{i}}{V_{i}^{2} V'_{i}}
\, (j'_i V_i - J_cV'_i)\, J^{(1)} . 
\label{A_1} 
\end{eqnarray}
In these equations the functions of the outer electric field profile are 
piecewise constant, for example, $V=V(E_1)\equiv V_1$ if $0<y<Y_1(s_1)$ and
$V=V(E_3)\equiv V_3$ if $Y_1(s_1)<y<1$. The subscripts $i$ are either 1 or 3
according to the value of $y$. The solutions that do not increase exponentially 
as $\tau\to - \infty$ (and thus they may match the previous stage) are 
\begin{eqnarray}
E_i^{(1)} = \frac{V_{i}}{J_{c} V'_{i}}\, J^{(1)}(\tau) - \frac{\mu_{i}\,
(j'_i V_i - J_cV'_i)}{J_{c} V'_{i} j'_{i}}\nonumber\\
\times \int_{-\infty}^{\tau} 
e^{-\mu_{i}\, (\tau - t)}\,  J^{(1)}(t)\, dt . 
\label{solE_1} 
\end{eqnarray}
Here 
\begin{eqnarray}
\mu_i = {(K_{i}+R_{i})j'_{i}\over V'_{i}}\, ,\quad i=1,3,\label{mu_i}
\end{eqnarray}
$i=1$ if $0<y<Y_1(s_1)$ and $i=3$ if $Y_1(s_1)<y<1$). The expression for 
$A^{(1)}$ immediately follows from (\ref{E_1}) and (\ref{solE_1}). 

    The equations at the injecting layer are (\ref{1bis}) and (\ref{3bis})
with $J=J_c + \epsilon\,\rho\, J^{(1)}(\tau) + O(\epsilon^2)$, while the 
boundary condition is $F(0,\tau) = E_2(J_c) + \epsilon\,\rho\, J^{(1)}(\tau) 
+ O(\epsilon^2)$. To determine $J^{(1)}(\tau)$ we shall substitute the outer 
field profile (\ref{outer2}) in the bias condition (\ref{4}). The  result is
\begin{eqnarray}
\phi = E_1(J_c) + [E_3(J_c) - E_1(J_c)]\, [1-Y_1(s_1)] \nonumber\\
+\epsilon\,\left\{ E^{(1)}_1 + [E^{(1)}_3 - E^{(1)}_1]\, [1-Y_1(s_1)] 
- [E_3(J_c)  \right.\nonumber\\
\left.    - E_1(J_c)] c_+ \tau
+ \int_0^\infty [F(x,\tau) - E_1(J_c)]\, dx \right.\nonumber\\
\left. + \int_{-\infty}^0 [F(\chi) - E_1(J_c)]\, d\chi \right.\nonumber\\
\left. + \int_0^\infty [F(\chi) - E_3(J_c)]\, d\chi \right\}  +
O(\epsilon^2). \label{bias2}
\end{eqnarray}
Let us define now 
\begin{eqnarray}
h(\tau) = [E_3(J_c) - E_1(J_c)]\, c_+ \tau  - \int_0^\infty [F(x,\tau) 
 \nonumber\\
- E_1(J_c)]\, dx - \int_{-\infty}^0 [F(\chi) - E_1(J_c)]\, d\chi\nonumber\\
 - \int_0^\infty [F(\chi) - E_3(J_c)]\, d\chi,   \label{hdef}
\end{eqnarray}
which is essentially the area lost by the motion of the old wavefront $Y_1$ 
during the time $\tau$ minus the excess area under the injecting layer. 
If we include in the leading order of the outer electric field the two terms 
of (\ref{bias2}) that represent the excess area in the transition layer at 
$Y_1$, and insert (\ref{solE_1}) in the result, we obtain
\begin{eqnarray}
\left[ {V_{1}Y_{1}(s_{1})\over J_{c} V'_{1}} +  {V_{3}[1-Y_{1}(s_{1})]\over 
J_{c}V'_{3}}\right] J^{(1)}(\tau) - \frac{\mu_{1}Y_{1}(s_{1})}{J_{c} V'_{1} 
j'_{1}}\, (j'_1 V_1  \nonumber\\
 - J_c V'_1)\, \int_{-\infty}^{\tau} e^{-\mu_{1} (\tau - t)}
\, J^{(1)}(t)\, dt - \frac{\mu_{3}\, [1-Y_{1}(s_{1})]}{J_{c} V'_{3} j'_{3}}
\nonumber\\
\times (j'_3 V_3 - 
J_cV'_3)\, \int_{-\infty}^{\tau} e^{-\mu_{3}\, (\tau - t)}\, J^{(1)}(t)\, 
dt = h(\tau) .   
\label{h}
\end{eqnarray}

After some elementary manipulations, we can transform this integral equation for
$J^{(1)}$ in a linear second order ordinary differential equation which
can be readily solved under the condition that the solution should not increase
as $\tau\to -\infty$. We obtain: 
\begin{eqnarray}
J^{(1)}(\tau) = \frac{h(\tau)}{a_{1}} + \int_0^\infty G(\sigma)\, \frac{h(\tau - 
\sigma)}{a_{1}}\, d\sigma\, , \label{J1}
\end{eqnarray}
\begin{eqnarray}
G(\sigma) = \frac{(\lambda_{1} + \mu_{1})(\lambda_{1} + \mu_{3}) 
e^{\lambda_{1}\sigma} - (\lambda_{2} + \mu_{1})(\lambda_{2} + \mu_{3}) 
e^{\lambda_{2}\sigma}}{\lambda_{1} - \lambda_{2}}\, .\label{G}
\end{eqnarray}
 Here $\lambda_1$ and $\lambda_{2}$ are the (negative) roots of the 
quadratic polynomial:
\begin{eqnarray}
a_{1}\lambda^2 + a_{2}\lambda + a_{3},    \label{lambda}
\end{eqnarray}
whose coefficients are
\begin{eqnarray}
a_{1} = {V_{1}Y_{1}(s_{1})\over J_{c} V'_{1}} +  {V_{3}
[1-Y_{1}(s_{1})]\over J_{c} V'_{3}}\, ,\label{a_1}\\
a_{2} = \mu_1 \, \left( {Y_{1}(s_{1})\over j'_{1}} 
+ {V_{3} [1-Y_{1}(s_{1})]\over J_{c} V'_{3}} \right) \nonumber\\
+ \mu_3 \, \left( {V_{1} Y_{1}(s_{1})\over
J_{c} V'_{1}} + {1-Y_{1}(s_{1})\over j'_{3}} \right)\, ,\label{a_2}\\
a_{3} = \mu_1 \mu_3 \, 
\left(\frac{Y_{1}(s_{1})}{j'_{1}} + \frac{1-Y_{1}(s_{1})}{j'_{3}}\right)\, . 
\label{a_3}
\end{eqnarray}
We can write (\ref{J1}) in another form that suggests a more transparent
interpretation:
\begin{eqnarray}
J^{(1)}(\tau) = J'(s_1)\, (\tau - \tau_0) - I(\tau) - \int_{0}^{\infty} 
G(\sigma) I(\tau-\sigma) d\sigma \nonumber\\
\equiv J'(s_1)\, (\tau - \tau_0) - \tilde{I}(\tau), \label{J1bis}
\end{eqnarray}
where $J'(s_1)$ is given by (\ref{a14}) at $J=J_c$, $G(\sigma)$ is the
kernel (\ref{G}), and
\begin{eqnarray}
I(\tau) = a_1^{-1}\, \int_0^{\infty} [F(x,\tau)-E_1(J_c)]\, dx, \label{I}\\
\tau_0 = \frac{1}{(E_{3}-E_{1}) c_{+}} \,\left\{\int_{-\infty}^{0} 
[F(\chi) - E_{1}(J_{c})]\, d\chi \right.   \nonumber\\
\left. + \int_{0}^{\infty} [F(\chi) - 
E_{3}(J_{c})]\, d\chi + \frac{J'(s_{1})}{(E_{3}-E_{1}) c_{+}}
\right.\nonumber\\
\left.  \times \left[ \frac{[1-Y_{1}(s_{1})] (j'_{3} V_{3}
- J_{c} V'_{3})}{J_{c}\mu_{3} V'_{3} j'_{3} } \right.\right.\nonumber\\
\left.\left. + \frac{Y_{1}(s_{1}) 
(j'_{1} V_{1}-J_{c} V'_{1})}{J_{c}\mu_{1} V'_{1} j'_{1}} \right] \right\} \, .
\label{tau_0}
\end{eqnarray}
The terms on the right side of (\ref{J1bis}) clearly display the balance 
between the area lost by the motion of the old wavefront at $Y_1(s)$ and 
the excess area under the injecting layer.

Let us recapitulate. We need to solve the following equation for the field
in the injecting layer:
\begin{eqnarray}
\frac{\partial^{2}{F}}{\partial{x} \partial \tau} + \frac{[J_c + \epsilon 
J^{(1)}(\tau)]\, [K(F)+R(F)]}{V(E)^{2}}\nonumber\\
\times \left(\frac{V'(F)}{K(F)+R(F)}\, 
\frac{\partial F}{\partial t} + V(F)\, \frac{\partial F}{\partial x} 
  \right. \nonumber\\ 
\left. + j(F) - J_c - \epsilon J^{(1)}(\tau)\right) = \epsilon\,\frac{1}{V(F)}\,
\frac{dJ^{(1)}}{d\tau}\,,    \label{Eq.contact}
\end{eqnarray}
or equivalently, the system
\begin{eqnarray}
\frac{\partial A}{\partial \tau} = {[J_c + \epsilon J^{(1)}(\tau)]\, 
[K(E)+R(E)]\over V(E)^{2}}\nonumber\\
\times [j(E) - V(E)\, A] , \label{sysa.contact} \\
\frac{\partial E}{\partial x} = {J_c + \epsilon J^{(1)}(\tau)\over V(E)} 
- A , \label{syse.contact}
\end{eqnarray}
for $A$ and $E=F(x,\tau)$, with the following boundary condition
\begin{eqnarray}
F(0,\tau) = \rho\, [J_c + \epsilon J^{(1)}(\tau)]. \label{bc.contact}
\end{eqnarray}
Here $J^{(1)}(\tau)$ is given by (\ref{J1}) as a functional of the function 
$h(\tau)$ or by (\ref{J1bis}) as a functional of the excess area under the
injecting layer, $I(\tau)$. Besides solving this problem, $F(x,\tau)$ needs 
to satisfy the matching condition: 
\begin{eqnarray}
F(x,\tau) - F_0(x;J(s)) \ll \epsilon, \label{a20}
\end{eqnarray}
as $s\to s_1-,\, \tau\to -\infty$ on an appropriate overlap domain. 
Here $F_0(x;J(s))$ is the 
quasistationary boundary layer solution of (\ref{a8}) and (\ref{a7}) 
for $s<s_1$, $J(s_1) = J_c$, with $J(s)$ given by (\ref{a14}). We have 
kept the $O(\epsilon)$ term in (\ref{bc.contact}) because it becomes of 
order 1 after the new wavefront(s) are formed. Eqs.~(\ref{Eq.contact}) 
- (\ref{a20}) constitute a meaningful problem which can be solved 
numerically or further studied by analytic means. Notice that the 
forcing term at the boundary, proportional to $J^{(1)}(\tau)$ is a 
functional of $h(\tau)$: the area lost as the old wavefront moves 
toward $y=1$ minus the area gained because of the field growth in the 
injecting layer. The present discussion could be used to make precise 
the observations of Higuera and Bonilla for the creation of a new wave 
in the usual model of the Gunn effect (see Section 5 of \cite{higPD92}). 
Anticipating the creation of new wavefronts near $y=0$, we 
decide to define their position on the slow space
scale by $Y_i = \epsilon X_i$ ($i=3,4$), $X_3<X_4$, such that
\begin{eqnarray}
F(X_3(\tau),\tau) = e_0 := {E_2(J_c) + E_3(J_c)\over 2}\, ,\label{a22}\\	
F(X_4(\tau),\tau) = E_0 := {E_1(J_c) + E_2(J_c)\over 2}\, .	
\label{a23}
\end{eqnarray}

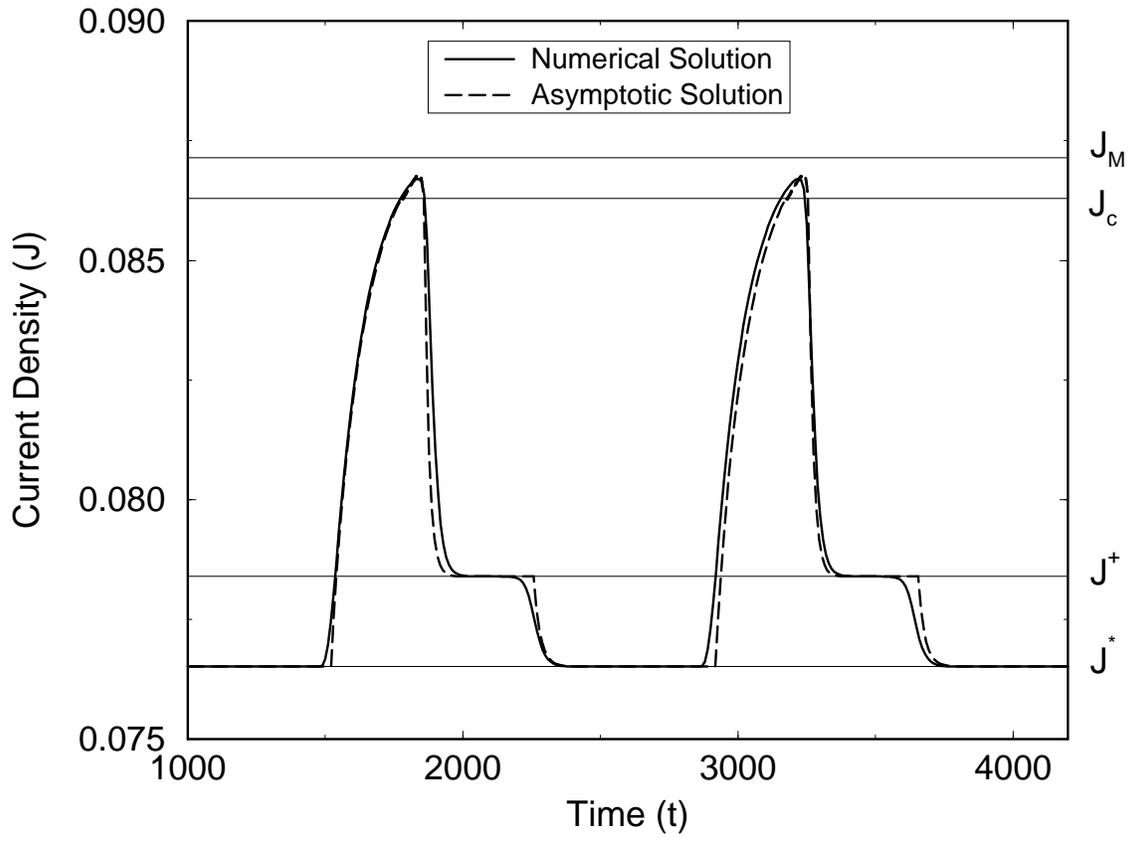
\begin{figure}   
\caption{ Comparison between the results of direct numerical simulation 
and the leading-order asymptotic approximation of the solution for the same
values of the parameters as in Figure 3.
}
\label{shedding}
\end{figure}

The results of our simulation are shown in Fig.~\ref{shedding}. When 
$F(x,\tau)$ is sufficiently close to $E_3[J(\tau)]$ on some space interval, 
we consider that the fast shedding stage has ended. Then the positions of 
the new wavefronts $Y_3$ and $Y_4$ and the value of the current density 
may be used as initial conditions for the following slow stage with 
three wavefronts as described in the previous Section. 

\subsection{A local limiting case when $\ln L\gg1$}
There is an asymptotic limit in which Eqs.~(\ref{Eq.contact}) -
(\ref{a20}) can be recast as a simple, local, universal problem.
 On the $\tau$ time scale and whenever $J^{(1)}(\tau)<0$, the 
quasistationary boundary layer solution of (\ref{a8}) and (\ref{a7}) 
for $s<s_1$, $J(s_1) = J_c$, with $J(s)$ given by (\ref{a14}) satisfies 
(see Appendix A of \cite{bonSJAM94}): 
\begin{eqnarray}
F_0(x;J(s)) \sim E_{2} - c_{L}\, \exp\left\{ - \frac{j'_{2}\, 
(x-X)}{V_{2}}\right\}
\label{Eleft}
\end{eqnarray}
[as $(x-X) \rightarrow	- \infty$], with
\begin{eqnarray}
c_{L} = (E_{2}-E_{0}) \exp\left\{ \int_{E_{0}}^{E_{2}} 
\left[\frac{1}{E-E_{2}}	\right.\right.\nonumber\\
\left.\left. - \frac{j'_{2}\, V(E)}{V_{2}\, [j(E)-J]} 
\right]\, dE \right\},	\label{cL}
\end{eqnarray}
and
\begin{eqnarray}
F_0(x;J(s)) \sim E_{1} + c_{R}\, \exp\left\{ - \frac{j'_{1}\, (x-X)}{V_{1}}
\right\} 	
\label{Eright}
\end{eqnarray}
[as $(x-X) \rightarrow	+ \infty$], with
\begin{eqnarray}
c_{R} = (E_{0}-E_{1}) \exp\left\{ - \int_{E_{1}}^{E_{0}} \left[
\frac{1}{E-E_{1}}	\right.\right.\nonumber\\
\left.\left. - \frac{j'_{1}\, V(E)}{V_{1}\, [j(E)-J]}
\right]\, dE \right\}.	\label{cR}
\end{eqnarray}
Here we have defined $X_4=X(s)$ such that $F_0(X(s);J(s))=E_0$ given by 
(\ref{a23}). By using the boundary condition at $x=0$, (\ref{J1bis}) and
the matching condition (\ref{a20}), together with (\ref{Eleft}), we obtain
\begin{eqnarray}
\epsilon\, \rho\, \left[J'(s_1)\, (\tau - \tau_0) - \tilde{I}(\tau)
\right] \sim - c_{L}\, e^{ \frac{j'_{2}X}{V_{2}}} ,
\label{X(s)}
\end{eqnarray}
which yields the instantaneous ``position'', $X(s)$, of the transition 
layer linking $E_2$ and $E_1$ as $s\to -s_1$. Of course during the
shedding stage the position of the transition layer is no longer given
by (\ref{X(s)}). To find it, we need a better approximation to the
field at the injecting contact than the quasistationary one. We shall
proceed as follows:
\begin{itemize}
\item We define (as before) $X(\tau)$ as the position for which the 
field in the injecting layer is $(E_1+E_2)/2$.
\item We find an approximate solution to Eqs.~(\ref{Eq.contact}) -
(\ref{a20}) in the limit $\Delta x\to -\infty$ where $x=X(\tau)+\Delta x$
and $1\ll |\Delta x|\ll X(\tau)$. This approximation breaks down if
$dX/d\tau = O(1)$, which occurs for $\tau = O(\ln\epsilon)$. This growth
of $dX/d\tau$ indicates that the end of the injecting layer starts advancing
and shedding of a new pulse starts.  
\item For those times, we can find a reduced shedding problem  with a
local boundary condition. Its solution describes the formation of a
wavefront joining $E_3(J_c)$ and $E_1(J_c)$ (which will eventually become
the foremost wavefront of the newly shed pulse, $Y_4(s)$, described in Section 
\ref{asymptotics}). After a long time, the back of the pulse is a 
stationary layer joining $E_2(J_c)$ and $E_3(J_c)$.
\item The formation of the back of the new pulse will be described by
an equation similar to the previous reduced shedding problem with 
different local boundary and matching conditions.
\end{itemize}

\subsubsection{Approximate form of the electric field near the injecting 
contact: forefront shedding} Far from $x=X(\tau)$ and close to $x=0$, we can 
linearize Eq.~(\ref{Eq.contact}) or equivalently, Eqs.~(\ref{sysa.contact}) 
and (\ref{syse.contact}) about $A=J_c/V(E_2(J_c))$ and $E=E_2(J_c)$. The 
result is 
\begin{eqnarray}
\frac{\partial^{2} \hat{E}}{\partial x\partial \tau} + {J_{c}\, [K_{2}+R_{2}]
\over V_{2}}\, \frac{\partial \hat{E}}{\partial x} +  {J_{c} V'_{2}
\over V_{2}^{2}}\, \frac{\partial \hat{E}}{\partial \tau}\nonumber\\
+ {J_{c}\, j'_{2}\, [K_{2}+R_{2}] \over V_{2}^{2}}\, \hat{E} 
= \epsilon \left( {1\over V_{2}}\, {dJ^{(1)}\over d\tau} \right.\nonumber\\
\left. + {J_{c}(K_{2}+R_{2})\over V_{2}^{2}}\, J^{(1)}\right)\, , 
\label{l1}\\
\hat{E}(0,\tau) = \epsilon\rho J^{(1)}(\tau). \label{l2}
\end{eqnarray}
We have ignored terms of the second order in $\epsilon$ and the linearized
field. The solution of Eqs.~(\ref{l1})-(\ref{l2}) which satisfies
\begin{eqnarray}
\hat{E}(x,\tau)\sim \epsilon\left[ \left(\rho - {1\over j'_{2}}\right) 
e^{-{j'_{2}x\over V_{2}}} + {1\over j'_{2}}\right]\, J^{(1)}(\tau) 
\label{l2match}
\end{eqnarray} 
(as $\tau\to - \infty$) is derived in Appendix \ref{appendix}. Its 
asymptotic behavior as $x \to +\infty$ corresponds to the transition
region where $x\approx X(\tau)\gg 1$:
\begin{eqnarray}
\hat{E} \sim \epsilon\, \left(\rho - {1\over j'_{2}}\right)\,
e^{- \frac{j'_{2}x}{V_{2}}}\, \left\{ J'(s_1)\, [\tau - \tau_0 - 
\theta\, x] \right.\nonumber\\
\left. - \tilde{I}(\tau - \theta x) \right\}\, ,\label{l3}\\
\theta = {J_{c} V'_{2} - j'_{2} V_{2} \over J_{c} V_{2} (K_{2}+R_{2})} > 0,
\label{l4}
\end{eqnarray}
To discuss the limit of validity of (\ref{l3}), we consider the region 
where $\hat{E} = O(1)$. For instance we may set 
\begin{eqnarray}
\epsilon\, 
\left(\rho - {1\over j'_{2}}\right)\, e^{- \frac{j'_{2}X}{V_{2}}}\,
[J'(s_1)\, (\tau - \theta X) - \tilde{I}(\tau - \theta X)] \nonumber\\
\sim - {E_{2} - E_{1}\over 2} ,\label{l7}
\end{eqnarray}
according to the definition of $X(\tau)$ as the position for which the
electric field takes on the value $(E_1+E_2)/2$. When $\tau<0$ and 
$|\tau|\gg x$ and using the fact that $|\ln\epsilon|\gg 1$, we just 
neglect the second term on the left side in (\ref{l7}) to obtain
\begin{eqnarray}
X(\tau)\sim {V_{2}\over |j'_{2}|}\,\ln\left(\frac{(E_{2} - E_{1}) j'_{2}}
{2 (1-\rho j'_{2}) J'(s_{1})\epsilon\tau} \right) \, . \label{l11}
\end{eqnarray}
This approximation (the {\em adiabatic approximation}) holds if $dX/d\tau\ll
1$. 
Our approximation continues to hold even when $\tau$ increases
to zero in such a way that $|\tau|\ll X$. Actually the adiabatic
approximation breaks down when the left side of (\ref{l7}) changes sign. 
Once this happens, we can no longer discard the term proportional to 
$\tilde{I}(\tau - \theta x)$ in (\ref{l7}). We have 
\begin{eqnarray}
\tilde{I}(\tau - \theta x) \sim K\, J'(s_{1})\, X(\tau - \theta X(\tau)),
\label{l8}\\
K = \frac{E_{2}-E_{1}}{(E_{3}-E_{1}) c_{+}}\, .\label{l9}
\end{eqnarray}
$K$ is approximately constant. Then the breakdown condition is
\begin{eqnarray}
\tau - \theta X(\tau) \sim {\tilde{I}(\tau - \theta x) \over J'(s_{1})}\sim
K\, X(\tau - \theta X(\tau)),
\label{l12}
\end{eqnarray}
whence
\begin{eqnarray}
\tau \sim (\theta + K)\, X(\tau) \sim {(\theta + K) V_{2}\over |j'_{2}|}
\nonumber \\ 
\times \ln \left( \frac{(E_{2} - E_{1}) (j'_{2})^{2}}{2 (1-\rho j'_{2}) V_{2} 
\theta J'(s_{1})\epsilon} \right)
\label{l13}
\end{eqnarray}
We now set new length and time scales as follows: 
\begin{eqnarray}
x = {V_{2}\over |j'_{2}|}\,\ln\left(\frac{(E_{2} - E_{1}) (j'_{2})^{2}}{2 
(1-\rho j'_{2}) V_{2} \theta J'(s_{1})\epsilon} \right) + \eta , \label{l14}\\
\tau = {(\theta + K) V_{2}\over |j'_{2}|}\, 
\ln\left(\frac{(E_{2} - E_{1}) (j'_{2})^{2}}{2 (1-\rho j'_{2}) V_{2} 
\theta J'(s_{1})\epsilon} \right) + \sigma .\label{l15}
\end{eqnarray}

While this change of coordinates leaves invariant (\ref{1bis}) and 
(\ref{3bis}), inserting condition (\ref{l7}) in (\ref{l4}), we find
\begin{eqnarray}
E(\eta,\sigma) \sim E_2(J_c) + {[E_{2}(J_c) - E_{1}(J_c)]\, |j'_{2}|\over 
2V_{2}\theta}\nonumber\\
\times (\sigma - \theta \eta)\, e^{{|j'_{2}|\eta\over V_{2}}}\, ,
\quad \mbox{as}\quad\eta\to -\infty,
\label{locbc.contact}
\end{eqnarray}

We are thus led to the following reduced shedding problem. Solve
for $-\infty<\eta<\infty$, and $-\infty<\sigma<\infty$:
\begin{eqnarray}
\frac{\partial A}{\partial \sigma} & = & {J_c \, [K(E)+R(E)]\over V(E)^{2}}\, 
[j(E) - V(E)\, A] , \label{loca.contact} \\
\frac{\partial E}{\partial \eta} & = & {J_c \over V(E)} - A , 
\label{loce.contact}
\end{eqnarray}
subject to the boundary condition (\ref{locbc.contact})
and to the matching condition (\ref{a20}) (once the changes of coordinates
(\ref{l14})-(\ref{l15}) are inserted in it). 

The boundary condition (\ref{locbc.contact}) is {\em local} and it 
represents a small wave whose amplitude increases exponentially as it  
moves to the right with velocity $1/\theta$. Notice that the same
interpretation can be made of the solution of the amplitude equation
describing the onset of the Gunn-type oscillations in the
supercritical case \cite{bonSST94}. Unlike the bifurcation case, the
solution of the problem (\ref{locbc.contact})-(\ref{loce.contact})
describes how the amplitude of the electric field wavefront grows until
it reaches $E=E_3(J_c)$, which happens at a time we will denote by 
$\tau^*(\epsilon)$. Then the injecting layer presents the following
structure: a forefront which is an advancing transition layer joining 
$E_3(J_c)$ and $E_1(J_c)$ and located at $X(\tau)$, a region of width 
$(\ln\epsilon)$ on which $E=E_3(J_c)$, and a quasistationary back front 
which is a transition layer joining $E_2(J_c)$ and $E_3(J_c)$.

\subsubsection{Approximate form of the electric field near the injecting 
contact: backfront shedding} 
Clearly what happens next, for $\tau>\tau^*$, is described by 
other equations. Here we shall only consider the stable case, $J_c>J^\dag$,
in which only one wave is shed. In particular the expression for 
$I(\tau)$ is different and $J^{(1)}(\tau)$ starts decreasing so as to keep 
the bias condition. There is another local problem which governs the 
transformation of the quasistationary backfront joining $E_2(J_c)$ and 
$E_3(J_c)$ into a moving wavefront joining $E_1(J_c)$ and $E_3(J_c)$. 
Since the necessary analysis is similar to that for the forefront, we 
will describe only its more salient features, omitting many details. 

Let us denote by $X_4(\tau)$ the instantaneous position of the forefront 
from now on, and let $\tilde{X}(\epsilon)\equiv X_4(\tau^*)$. Let us
make a new change of coordinates which is simply a shift of 
(\ref{l14})-(\ref{l15}):
\begin{eqnarray}
x = \tilde{X}(\epsilon) + \eta , \label{l16}\\
\tau = \tau^*(\epsilon) + \sigma .\label{l17}
\end{eqnarray}

The excess area in the injecting boundary layer is no longer 
proportional to $(E_2-E_1)\, X_4$. We have instead that for $\tau>\tau^*$,
this area is equal to the sum of: (i) the excess area at the time 
$\tau^*$, (ii) the excess area due to the motion of the forefront between
the times $\tau^*$ and $\tau$, minus the loss of area due to the motion
of the backfront located at $X_3(\sigma)$: 
\begin{eqnarray}
a_1\, I(\sigma)\equiv \int_{0}^\infty [F(x,\sigma) -E_1(J_c)] \, dx =
 (E_2 - E_1) \tilde{X}   \nonumber\\
+ (E_3 - E_1)\, c_-\, \sigma - (E_3 - E_2) \, [\tilde{X} - X_3(\sigma)]  
\label{excess}
\end{eqnarray}
$X_3(\sigma)$ is defined as the position for which the electric field takes 
on the value $(E_3+E_2)/2$. Then 
\begin{eqnarray}
\epsilon\, 
\left(\rho - {1\over j'_{2}}\right)\, e^{- \frac{j'_{2}X_{3}}{V_{2}}}\,
[J'(s_1)\, (\tau - \theta X_{3}) \nonumber\\
- \tilde{I}(\tau^* + \sigma - \theta X_{3})] 
\sim {E_{3} - E_{2}\over 2} ,\label{eq:X_3}
\end{eqnarray}
and therefore we can make an approximation similar to (\ref{l11}),
\begin{eqnarray}
X_3(\sigma)\sim {V_{2}\over |j'_{2}|}\,\ln\left(\frac{\Gamma}{\epsilon\sigma} 
\right) \sim \tilde{X} - {V_{2}\over |j'_{2}|}\,\ln\sigma\, , \label{X_3}
\end{eqnarray}
for times $\sigma < \theta\tilde{X}$. $\Gamma>0$ is a constant which we
shall leave unspecified. After times $\sigma=\theta\tilde{X}$, the
approximation given by (\ref{l8}) breaks down and $\tilde{I}$ in 
(\ref{eq:X_3}) should be replaced by (\ref{excess}). 
The adiabatic approximation breaks down when
the left side of (\ref{eq:X_3}) changes sign, which yields new times
$\sigma^*$ and positions $\tilde{X}_3$, after which the backfront is
shed. To analyze the shedding problem we define new variables $\sigma$
and $\eta$ by using (\ref{l16}) and (\ref{l17}) with $\sigma^*$ and 
$\tilde{X}_3$ instead of $\tau^*$ and $\tilde{X}$. The result is that
the shedding of the backfront is governed by Eqs.~(\ref{loca.contact}) - 
(\ref{loce.contact}) with the following boundary condition instead of 
(\ref{locbc.contact}): 
\begin{eqnarray}
E(\eta,\sigma) \sim E_2(J_c) - {E_{3} - E_{2}\over 2\tau'}\,
(\sigma - \theta \eta)\, e^{{|j'_{2}|\eta\over V_{2}}}
\label{locbc.back}
\end{eqnarray}
(as $\eta\to -\infty$; $\tau'$ is a positive constant), and the new 
matching condition with the coresponding stationary state.  

\section{Conclusions}
\label{conclusion}
We have shown that a unipolar drift-diffusion model of charge transport
in ultrapure p-Ge has solutions existing globally in time. We have
performed a new asymptotic analysis of a reduced version of this model 
which describes the trap-dominated slow Gunn effect on a long sample. The 
building blocks of this analysis are the heteroclinic orbits used to 
construct the typical solitary waves mediating Gunn-like oscillations. 
During most of the oscillation, the motion of the heteroclinic orbits
and the change of the electric field inside and outside the solitary
waves (enclosed by heteroclinic orbits) follow adiabatically the 
evolution of the total current density. When a solitary wave reaches
the receiving contact, the current increases until the injecting
contact layer becomes unstable and sheds new wave(s). The wave shedding 
stage of the oscillation is described by a semi-infinite problem
which needs to be solved and matched to the rest of the oscillation.
As an outcome, we have found a criterion that shows that single or
multiple wave shedding is possible during each oscillation, depending
on the resistivity of the injecting contact. While single shedding is
the usual (stable) Gunn effect, multiple wave shedding breaks the spatial 
coherence of the sample. This new instability mechanism may eventually 
explain the complicated behavior observed in experiments performed in long 
semiconductor samples \cite{kahn45PRB92,kahnPRB92,kahnPRL92}.
We have confirmed these results by direct numerical simulation of the
reduced model, in particular the new predictions of multiple shedding
of solitary waves in the unstable case. A simpler asymptotic description 
is possible in the limit $\ln L\to\infty$: we need to solve two spatially
infinite problems, one for the shedding of the forefront of the new
solitary wave, the other for its backfront. Although this limit is
unrealistic and extremely hard to compare with numerical simulations
($L$ should be so large that the needed computing time is excessive),
we have analyzed it in some detail in view of the insight offered in
the shedding problem. 

We expect that a similar analysis of the Gunn effect can be performed
in other models, although the details of the shedding problem will in
general be different \cite{bon96}.

\section{ACKNOWLEDGMENTS}
\label{acknowledgements}

It is a pleasure to acknowledge beneficial conversations with S.\ Venakides.  
We acknowledge financial support from the the Spanish DGICYT
through grants PB92-0248, PB93-0438 and PB94-0375, the NATO Travel 
Grant Program through grant CRG-900284, and the EC Human Capital 
and Mobility Programme under contract ERBCHRXCT930413.

\appendix
\setcounter{equation}{0}
\section{Solution of the problem (73)-(75)}
\label{appendix}
In the sequel we shall study the form and asymptotic properties of the
solutions of (\ref{l1})-(\ref{l2match}). Eq.~(\ref{l1}) may be written as
\begin{eqnarray}
{\cal{L}} \hat{E}\equiv \frac{\partial^{2} \hat{E}}{\partial x\partial \tau} 
+ D\,\frac{\partial\hat{E}}{\partial x} + B\,\frac{\partial \hat{E}}{\partial 
\tau} - C  \hat{E} \nonumber\\
=   {\epsilon\over V_{2}}\, \left( {dJ^{(1)}\over 
d\tau} + D\, J^{(1)}\right)\, , \label{A1}\\
B = {J_{c} V'_{2}\over V_{2}^{2}}\,,\quad\quad
C = -{J_{c}\, j'_{2}\,[K_{2}+R_{2}]\over V_{2}^{2}} = - B\mu_2 , \nonumber\\
D = {J_{c}(K_{2}+R_{2})\over V_{2}} =  {B\mu_{2} V_{2}\over j'_{2}}\,.
\label{A2}
\end{eqnarray}
We now write $\hat{E}(x,\tau)$ as a sum of the solutions of simpler problems: 
\begin{eqnarray}
\hat{E}(x,\tau) = \epsilon \sum_{i=1}^{4} {\cal{E}}_i(x,\tau),
\label{A3}\\
{\cal{L}}{\cal{E}}_1 = 0,\quad\quad {\cal{E}}_1(0,\tau) = \rho J'(s_1) (\tau - 
\tau_0),\label{A3.1}\\
{\cal{L}}{\cal{E}}_2 = {J'(s_{1})[1+D(\tau -\tau_{0})]\over V_{2}}\,,\quad\quad 
{\cal{E}}_2(0,\tau) = 0,\label{A3.2}\\
{\cal{L}}{\cal{E}}_3 = 0,\quad\quad {\cal{E}}_3(0,\tau) = -\rho \tilde{I}(\tau), 
\label{A3.3}\\
{\cal{L}}{\cal{E}}_4 = -{1\over V_{2}}\, \left( {d\tilde{I}\over d\tau} + D
\tilde{I}\right) \equiv g(\tau)\,,\quad\quad {\cal{E}}_4(0,\tau) = 0.
\label{A3.4}
\end{eqnarray}
The solutions of (\ref{A3.1}) and (\ref{A3.2}) are immediate:
\begin{eqnarray}
{\cal{E}}_1 = \rho\, J'(s_1)\, (\tau - \tau_0 - \theta\, x)\, e^{\frac{Cx}{D}} ,
\label{A4.1}\\
{\cal{E}}_2 = {D J'(s_{1})\over C V_{2}}\, [(\tau - \tau_0 - \theta\, x)\, 
e^{\frac{Cx}{D}} - (\tau - \tau_0)],
\label{A4.2}
\end{eqnarray}
where $\theta$ is given by (\ref{l4}). To calculate ${\cal{E}}_3$, we observe 
that, on taking Fourier transform in (\ref{A3.3}), we obtain 
\begin{eqnarray}
{\cal{E}}_3(x,\tau) = - \rho \int_{-\infty}^\tau K(x,\tau - s)\, \tilde{I}(s)\, 
ds , \label{A5}
\end{eqnarray}
where 
\begin{eqnarray}
K(x,\tau) = {1\over 2\pi}\, \int_{-\infty}^\infty \tilde{K}(x,\omega)\, 
e^{i\omega\tau}\, d\omega , \label{A6.1}\\
\tilde{K}(x,\omega) =  e^{-Bx} + e^{-Bx}\left[\exp\left( \frac{(D B + C)x}
{D + i\omega}\right) - 1\right]\,.   \label{A6.2}
\end{eqnarray}
Hence 
\begin{eqnarray}
K(x,\tau) = e^{-Bx}\left\{ \delta(\tau) + {1\over 2\pi}\, \int_{-\infty}^\infty 
\left[ e^{ \frac{(D B + C)x}{D + i\omega}} \right.\right.
\nonumber\\
\left.\left. - 1\right]\,
e^{i\omega\tau}\, d\omega\right\}
\equiv I_1(x,\tau) +  I_2(x,\tau), \label{A7}
\end{eqnarray}

The term $I_2(x,\tau)$ in (\ref{A7}) can be evaluated by standard complex
variable methods. To this end, we notice that the residue of the function 
exp$i(\zeta\tau-M/\zeta)$ at $\zeta=0$ is 
$${1\over i}\,\sum_{l=0}^{\infty} {M^{l+1}\tau^{l}\over l! (l+1)!}\, .$$
Then Cauchy's theorem yields
\begin{eqnarray}
I_2(x,\tau) = e^{-Bx - D\tau} \, (DB + C)\, x\nonumber\\
\times \Psi[(DB + C)x\tau]\, \eta(\tau) ,
\label{A8}
\end{eqnarray}
where $\eta(\tau) = 1$ for $\tau>0$ and $\eta(\tau) = 0$ otherwise. The 
function $\Psi$ is defined as follows
\begin{eqnarray}
\Psi(z) = \sum_{l=0}^{\infty} {z^{l}\over l! (l+1)!} \equiv {I_{1}(2\sqrt{z})
\over \sqrt{z}}\,,
\label{A9}
\end{eqnarray}
where $I_{1}(z)$ is the standard modified Bessel function of the
first kind with index 1. From (\ref{A5})-(\ref{A9}), we obtain
\begin{eqnarray}
{\cal{E}}_3(x,\tau) = - \rho\, e^{-Bx}\,\left\{ \tilde{I}(\tau) + (DB + C)\, 
x \right.\nonumber\\
\left. \times\, \int_{-\infty}^\tau e^{-D(\tau-s)} \Psi[(DB + C)\, x\, 
(\tau-s)]\, \tilde{I}(s)\, ds \right\} . \label{A10}
\end{eqnarray}

We now turn our attention to ${\cal{E}}_4$. We claim that 
\begin{eqnarray}
{\cal{E}}_4(x,\tau) = F(\tau) - \int_{-\infty}^\tau K(x,\tau-s)\, F(s)\, 
ds , \label{A11}
\end{eqnarray}
where $F(\tau)$ is any arbitrary solution of the equation
\begin{eqnarray}
B\, {dF\over d\tau} - C F= g(\tau), \label{A.F}
\end{eqnarray}
with $g(\tau)$ as in (\ref{A3.4}). 
At this juncture the reader may wonder whether (i) ${\cal{E}}_4$ thus
defined is actually independent of the choice of $F$, and (ii) 
${\cal{E}}_4$ is such that ${\cal{E}}_4(x,\tau)$ depends only on those 
values of $g(s)$ for which $s\le \tau$. It turns out that the answer
to these questions is yes. However, not to interrupt the main flow of the 
arguments  here, we shall postpone the discussion of this point till the
last part of this Appendix. We continue by describing in some detail the
asymptotics of ${\cal{E}}_3(x,\tau)$ and ${\cal{E}}_4(x,\tau)$ as $x\to\infty$. 
This is all we need to get the asymptotic form of $\hat{E}$ in view of the 
explicit form of ${\cal{E}}_1$ and ${\cal{E}}_2$ given by (\ref{A3.1}) and 
(\ref{A3.2}), respectively. 

Let us begin by ${\cal{E}}_3$. Since by (\ref{l4}) and (\ref{A2}) $\theta = 
(B+C/D)/D$, we need to study the behavior of the second term in the right side
of (\ref{A10}):
\begin{eqnarray}
R(x,\tau) = - \rho D^2 \theta x e^{-Bx}\,\int_{-\infty}^\tau e^{-D(\tau-s)}
\nonumber\\
\times \Psi[D^2 \theta x (\tau-s)]\, \tilde{I}(s)\, ds , \label{A12}
\end{eqnarray}
as $x\to\infty$. Without loss of generality, we may assume that 
$x (\tau-s)\gg 1$ and make use of the approximation $I_1(z)\sim (2\pi z)^{-1/2}
e^z$, as $z\to\infty$. Then 
\begin{eqnarray}
R(x,\tau) \sim  - {\rho \theta x e^{-Bx}\over 2\sqrt{\pi}}\,
\int_{-\infty}^\tau e^{-D(\tau-s)}\,
\frac{e^{2D\sqrt{\theta x (\tau-s)}}}{[D^{2}\theta x (\tau-s)]^{{3\over 4}}} 
 \nonumber\\
\times \tilde{I}(s)\, ds - {\rho\over 2}\, \sqrt{{\theta x\over \pi D^{3}}}\, 
e^{-Bx} \int_0^\infty \tilde{I}(\tau-\theta x\xi)\,\xi^{-{3\over 4}}
 \nonumber\\
\times e^{D\theta x(2\sqrt{\xi} -\xi)}\, d\xi . \label{A13}
\end{eqnarray}

In view of its definition (\ref{J1bis}) and (\ref{I}), it is natural to 
assume that $\tilde{I}(\tau)$ has algebraic behavior as $\tau\to\infty$.
Then the main contribution to $R(x,\tau)$ will correspond to $\xi\sim 1$,
and we may approximate the integral by Laplace's method, thereby obtaining 
\begin{eqnarray}
{\cal{E}}_3(x,\tau)\sim R(x,\tau) \sim - \rho\, e^{{Cx\over D}}\,
\tilde{I}(\tau -\theta x)  \label{A14}
\end{eqnarray}
(as $x\gg 1$), after substituting the result in (\ref{A10}). Let us consider 
now ${\cal{E}}_4(x,\tau)$ given by (\ref{A11}). By selecting the function $F$  
having a zero at $\tau$, $F(\tau)=0$, we observe that
\begin{eqnarray}
{\cal{E}}_4(x,\tau) = - \int_{-\infty}^\tau K(x,\tau-s) \, F(s)\, ds
\nonumber\\
\sim - {D\over C V_{2}}\, e^{{Cx\over D}}\,\tilde{I}(\tau -\theta x), 
\quad\mbox{as}\quad x\gg 1. \label{A15}
\end{eqnarray}

Putting together (\ref{A4.1}), (\ref{A4.2}), (\ref{A14}) and (\ref{A15}),  
we finally get 
\begin{eqnarray}
\hat{E}(x,\tau) \sim \epsilon\, \left(\rho + {D\over C}\right)\, 
e^{\frac{Cx}{D}}\,\left[ J'(s_1)\, (\tau - \tau_0 - \theta\, x) 
\right.\nonumber\\
\left. - \tilde{I}(\tau - \theta x) \right]\, ,\label{A16}
\end{eqnarray}
as $x\gg 1$. It is easy to check that this equation is the same as 
(\ref{l3}) by using the definitions (\ref{A2}).

We conclude this Appendix by stressing that ${\cal{E}}_4(x,\tau)$ given by
(\ref{A11}) is indeed well defined. This will be consequence of the 
following Liouville-type result, which is of some independent interest:
Let $u(x,\tau)$ be a solution of
\begin{eqnarray}
\frac{\partial^{2} u}{\partial x\partial \tau} + D\, {\partial u\over
\partial x} + B\,\frac{\partial u}{\partial \tau} - C  u = 0,
\label{A17}
\end{eqnarray}
where $D$, $B$ and $C$ are positive constants, and assume that $u$
is such that
\begin{eqnarray}
u(0,\tau) = 0,\nonumber\\
 |u(\cdot,\tau)|\le \nu(x)\, e^{\sigma\, |\tau|} ,
\quad \mbox{as}\quad\tau\to\infty \quad\mbox{with}\quad \sigma < D.
\label{A18}
\end{eqnarray}
Then $u(x,\tau)\equiv 0$. To check this, we set $\tau = -s$ in (\ref{A17})
and take the Laplace transform of this equation, thereby obtaining
\begin{eqnarray}
(D-z)\, {\partial U\over\partial x} - (Bz + C)\, U = - {\partial u(x,0)
\over\partial x} \nonumber\\
- B u(x,0) ,    \label{A19}
\end{eqnarray}
for $U(x,z)$, the Laplace transform of $u(x,\tau)$. Assuming that 
(\ref{A18}) holds, $U$ will be analytic in Im$z>0$, and integration of 
(\ref{A19}) yields 
\begin{eqnarray}
U(x,z) = {1\over D-z}\, \int_0^x \exp\left[{Bz + C\over D-z}\, (x-\xi)\right]
\nonumber\\
\times \left(- {\partial u(\xi,0)\over\partial x} - B u(\xi,0)\right)\, d\xi .
    \label{A20}
\end{eqnarray}
Then $U(x,z)\to 0$ as $|z|\to\infty$, and therefore $U(x,z)\equiv 0$,
whence the result. To derive the desired result for ${\cal{E}}_4(x,\tau)$,
we merely observe that
\begin{eqnarray}
\int_{-\infty}^\tau K(x,\tau-s)\, e^{{C\over B}s}\, ds
= e^{{C\over B}\tau}. \label{A21}
\end{eqnarray}
Indeed the function on the left-hand side above is a solution of
(\ref{A17}), and so is $e^{{C\over B}\tau}$. Since the difference
between these two solutions is at most algebraic as $\tau\to - \infty$,
(\ref{A18}) holds  and the result follows.

Using the fact that two different solutions of (\ref{A.F}) differ by a
term $k_1 \, e^{C\tau/B}$ (for some real constant $k_1$), it follows
from (\ref{A11}) and (\ref{A21}) that ${\cal{E}}_4$ is independent of
the particular choice of $F$. On the other hand, if we select $F$
(for a given $\tau=\bar{\tau}$) as follows
\begin{eqnarray}
F(s;\bar{\tau}) = \int_{s}^{\bar{\tau}} e^{{C\over B}(\bar{\tau}-\xi)}\, 
g(\xi)\, d\xi, \label{A22}
\end{eqnarray}
then it turns out that for $s<\bar{\tau}$, $F$ only depends on the
values of $f(\xi)$ with $\xi<\bar{\tau}$. By (\ref{A11}) the second claim
on ${\cal{E}}_4(x,\tau)$ is thus proved.

\bibliographystyle{prsty}

\begin{thebibliography}{10}

\bibitem[*]{bonilla:email} {E-address \tt bonilla@ing.uc3m.es}. Fax:
34-1-6249445. Author to whom all correspondence should be addressed.

\bibitem[\dag]{pjh:email} {E-address \tt pedroj@math.uc3m.es}.

\bibitem[\ddag]{herrero:email} {E-address \tt herrero@sunma4.mat.ucm.es}.

\bibitem[\S]{kindelan:email} {E-address \tt kindelan@dulcinea.uc3m.es}.

\bibitem[\S\S]{juanjo:email} {E-address \tt velazque@sunma4.mat.ucm.es}.

\bibitem{teitsPRL83} S.~W. Teitsworth, R.~M. Westervelt, and E.~E. Haller,
Phys. Rev. Lett. {\bf 51}, 825 (1983).

\bibitem{kahnPRB91} A.~M. Kahn, D.~J. Mar, and R.~M. Westervelt, Phys. Rev. B
{\bf 43}, 9740 (1991).

\bibitem{gunIJRD64} J.~B. Gunn, IBM J.\ Res.\ Dev.\ {\bf 8}, 141 (1964).

\bibitem{kahnPRB92} A.~M. Kahn, D.~J. Mar, and R.~M. Westervelt, Phys. Rev. B
{\bf 46}, 7469 (1992).

\bibitem{kahn45PRB92} A.~M. Kahn, D.~J. Mar, and R.~M. Westervelt,
Phys. Rev. B {\bf 45}, 8342 (1992).

\bibitem{kahnPRL92} A.~M. Kahn, D.~J. Mar, and R.~M. Westervelt,
Phys. Rev. Lett. {\bf 46}, 369 (1992).

\bibitem{teitsPRL84} S.~W. Teitsworth and R.~M. Westervelt,
Phys. Rev. Lett. {\bf 53}, 2587 (1984).

\bibitem{wesJAP85} R.~M. Westervelt and S.~W. Teitsworth, J. Appl. Phys. {\bf
57}, 5457 (1985).

\bibitem{bonPRB92} L.~L. Bonilla, Phys. Rev. B {\bf 45}, 11642 (1992).

\bibitem{haeIP89} N.~M. Haegel and A.~M. White, Infrared Phys. {\bf 29}, 915 (1989).

\bibitem{kro68} H. Kroemer, IEEE Trans. {\bf ED-15}, 819 (1968).

\bibitem{bonSST94} L.~L. Bonilla, I.~R. Cantalapiedra, M.~J. Bergmann, and
S.~W. Teitsworth, Semicond.\ Sci.\ Technol.\ {\bf 9}, 599 (1994).

\bibitem{mitAPA86} V.~V. Mitin, Appl. Phys. A {\bf 39}, 123 (1986).

\bibitem{schSSE88} E. Sch\"{o}ll, Solid-State Electron.\ {\bf 31}, 539 (1988).

\bibitem{schAPA89} E. Sch\"{o}ll, Appl. Phys. A {\bf 48}, 95 (1989).

\bibitem{kuhPRB93} T. Kuhn {\it et~al.}, Phys. Rev. B {\bf 48}, 1478 (1993).

\bibitem{canPRB93} I.~R. Cantalapiedra, L.~L. Bonilla, M.~J. Bergmann, and
S.~W. Teitsworth, Phys.  Rev. B {\bf 48}, 12278 (1993).

\bibitem{bergPRB96} M.~J. Bergmann, S.~W. Teitsworth, L.~L. Bonilla and I.~R. 
Cantalapiedra, Phys. Rev. B {\bf 53}, 1327 (1996).

\bibitem{teits94} S.~W. Teitsworth, M.~J. Bergmann, and L.~L. Bonilla, in {\em
Nonlinear Dynamics and Pattern Formation in Semiconductors and Devices},
Vol.~79 of {\em Springer Proceedings in Physics}, edited by
F.-J. Niedernostheide (Springer-Verlag, Berlin-Heidelberg, 1995), pp.\ 46--69.

\bibitem{bonPD91} L.~L. Bonilla and S.~W. Teitsworth, Physica D {\bf 50}, 545
(1991).

\bibitem{bonPD92} L.~L. Bonilla, Physica D {\bf 55}, 182 (1992).

\bibitem{henry}
D. Henry, {\em Geometric theory of semilinear equations}, Vol.~840 of {\em 
Lecture Notes in Mathematics} (Springer-Verlag, Berlin-Heidelberg, 1981).

\bibitem{bonSJAM94} L.~L. Bonilla, F.~J. Higuera, and S. Venakides, SIAM J.\
Appl.\ Math.\ {\bf 54}, 1521 (1994).

\bibitem{higPD92} F.~J. Higuera, and L.~L. Bonilla, Physica D {\bf 57}, 161 
(1992).

\bibitem{bon96}
L.~L. Bonilla and I.~R. Cantalapiedra, Preprint, 1996.

\end{thebibliography}

\newpage
\clearpage

\setcounter{figure}{0}

\begin{figure}
\begin{center}
\epsfig{file=fig1.eps, width=12cm, angle=-90}
\label{fig:1}
\vspace{1cm}
\caption[]{} 
\end{center}
\end{figure}

\newpage
\clearpage
\begin{figure}
\begin{center}
\epsfig{file=fig2.eps, width=12cm, angle=-90}
\label{fig:2}
\vspace{1cm}
\caption[]{} 
\end{center}
\end{figure}

\newpage
\clearpage
\begin{figure}
\begin{center}
\epsfig{file=fig3.eps, width=11cm, angle=-90}
\label{fig:3}
\vspace{1cm}
\caption[]{} 
\end{center}
\end{figure}

\newpage
\clearpage
\begin{figure}
\begin{center}
\epsfig{file=fig4.eps, width=11cm, angle=-90}
\label{fig:4}
\vspace{1cm}
\caption[]{} 
\end{center}
\end{figure}

\newpage
\clearpage
\begin{figure}
\begin{center}
\epsfig{file=fig5.eps, width=11cm, angle=-90}
\label{fig:5}
\vspace{1cm}
\caption[]{} 
\end{center}
\end{figure}

\end{document}